\documentclass{aastex63}

\usepackage{bm}
\usepackage{mathrsfs}
\usepackage{amsbsy}
\usepackage{amsmath}

\newcommand{\Linit}{L_{\rm init}}

\newcommand{\msun}{M$_{\rm \odot}$~}

\newcommand{\bcrit}{b_{\rm crit}}
\newcommand{\vinf}{v_{\infty}}
\newcommand{\chieff}{$\chi_{\mathrm{eff}}$}

\shorttitle{GW capture with spin}
\shortauthors{Bae et al.}

\graphicspath{{./}{figures/}}

\begin{document}

\title{Gravitational Wave Capture in Spinning Black Hole Encounters}

\correspondingauthor{Gungwon Kang}
\email{gwkang@kisti.re.kr}

\author{Yeong-Bok Bae}
\affiliation{National Institute for Mathematical Sciences, 70, Yuseong-daero 1689 beon-gil, Yuseong-gu, Daejeon, 34047, Korea}
\affiliation{Astronomy Program Department of Physics and Astronomy, Seoul National University, 1 Gwanak-ro, Gwanak-gu, Seoul  08826, Korea}

\author{Hyung Mok Lee}
\affiliation{Korea Astronomy and Space Science Institute, 776 Daedeokdae-ro, Yuseong-gu, Daejeon 34055, Korea}
\affiliation{Astronomy Program Department of Physics and Astronomy, Seoul National University, 1 Gwanak-ro, Gwanak-gu, Seoul  08826, Korea}

\author{Gungwon Kang}
\affiliation{Supercomputing Center at KISTI, 245 Daehak-ro, Yuseong-gu, Daejeon 34141, Korea}


\begin{abstract}

The orbits of two black holes which are initially unbound can be transformed into bound orbits by emitting gravitational waves during close encounters in a star cluster, which is called a gravitational wave (GW) capture. The effects of spin of black holes on GW capture are investigated in the context of numerical relativity. The radiated energy during the encounter is dependent on the effective spin when the black holes have the equal masses as expected from post-Newtonian approximation. The strongest emission is produced when the spins of both black holes are anti-aligned to the orbital angular momentum in the case of fly-by encounters. But the opposite is true in the case of direct merging: the strongest emission comes from the black holes with aligned spins to the orbital angular momentum. The fraction of direct merging among the GW captures increases in proportional to $v^{4/7}$ assuming the uniform distribution of pericenter distances in the encounters, where $v$ is the velocity dispersion of cluster, which means about 5\% of GW capture leads to the direct merging for star clusters with $v=150$ km s$^{-1}$.

\end{abstract}



\section{Introduction} \label{sec:intro}

Since the first detection of gravitational waves (GWs) by LIGO \citep{gw150914} in 2015, several more black hole (BH) binaries \citep{gw151012,gw151226,gw170104,gw170608,gw170814,gw170729etc} and a neutron star binary \citep{gw170817} have been discovered during the first and second observing runs of LIGO and Virgo. The BH binaries can be formed through the dynamical processes such as three body process \citep[e.g.,][]{bae14,park17} or gravitational radiation driven capture \citep{quinlan87,quinlan89,hansen14,bae17}, as well as from the evolution of stellar binary \citep{postnov14}. In the central region of the star cluster like a globular cluster or a nuclear star cluster in galaxies, significant number of binaries can be formed through the dynamical processes.

The gravitational radiation driven capture, or also known as gravitational wave capture (GW capture) takes place by losing the orbital energy through the emission of the GWs during close encounter between unbound BHs. GW capture is generally considered to be effective in the galactic nucleus environments rather than in the globular clusters because it requires a higher density of the BHs \citep{hong15,rasskazov19}. But recent studies including post-Newtonian (PN) dynamics reveal that GW capture can occur frequently in chaotic binary-single, or binary-binary interactions \citep{samsing14,samsing17,samsing18,samsing_askar18,rodriguez18,samsing19,zevin19}, even in the globular clusters.

One of the interesting consequences of the GW capture is that it can generate very eccentric compact binaries which can merge with significant eccentricity. Some fraction of BH binaries formed by GW capture do not have enough time to be circularized by emitting GWs before the merger. They can have significant eccentricities when they pass the frequency band of the present interferometric GW detector or even they merge. Thus their gravitational waveforms can be different from those of circular orbit. Depending on the orbits, they can have modulated waveforms due to the eccentricity or repeated burst-like waveforms.

Besides the GW capture of traditional astrophysical BHs in the central region of the cluster, several studies have suggested that the GW capture between primordial BHs \citep{bird16,mandic16,raidal17,sasaki18} could also take place. The BH masses with 20--100 \msun still remain open window \citep{bird16} for the dark matter, so that the primordial BHs in that range might be the constituents of the dark matter, and close encounters between them can lead to the gravitationally bound states by emitting GWs.

In previous paper \citep[][hereafter Paper I]{bae17} and this paper, we have focused on the GW capture with very close encounters. Several studies have already investigated the GW capture or the properties of the hyperbolic orbit by using PN calculations \citep{hansen72,quinlan87,quinlan89,mouri02,cho18}, but very close encounters including merger require full general relativistic approach. We have performed the numerical relativistic (NR) simulations, and compared them with the PN results.

The GW capture in full general relativity had been studied previously for equal mass BHs \citep{hansen14} and unequal mass ones \citep{bae17}. The critical impact parameter ($\bcrit$) causing the marginal capture of BHs depends on the relative velocity at infinite distance ($\vinf$) as $\bcrit\propto\vinf^{-9/7}$ in weak encounters, which can be derived by the 2.5 PN calculations \citep{hong15,bae17}. However, as the encounter becomes stronger---short pericenter with high relative velocity, the $\bcrit$ deviates from the PN results and becomes larger. The deviation from PN is more conspicuous for the encounters with high mass ratio. Thus the PN calculations of GW capture should be used carefully for the strong encounters with high mass ratios.

BH spins had not been considered in Paper I, but it is well known that spins can affect the orbits as well as the GW radiations. Many studies including \citet{campanelli06} have reported the changes of the BH orbits caused by the spin directions in the quasi-circular orbit. In this study, we have investigated the effects of BH spins on the GW radiations and the impact parameters in GW capture with the NR simulations.

In the following Section \ref{sec:setup}, the assumptions of our study and the setup of general relativistic simulation are described. In Section \ref{sec:results} the radiated energies depending on the spin directions, magnitudes and orbital angular momenta are presented, and the critical impact parameters for marginal capture are calculated from the radiated energy of the parabolic fly-by orbits. In addition, the results of NR simulations are compared with those of PN results and their discrepancies are presented. In Section \ref{sec:discuss} we discuss several noticeable points in our study and the astrophysical applications. In Appendix, the results of convergence tests with different resolutions are presented.


\section{Assumptions and Numerical setup} \label{sec:setup}

In order to simulate the marginally capturing hyperbolic orbit which gives the cross section for GW capture to take place, we should know its initial energy and angular momentum in advance. Since we don't know those values beforehand, we have to run it over and over again to find it out. For instance, we can find the initial condition of the marginally capturing orbit for a given initial angular momentum by changing the initial orbital energy. However, this method is very consumptive because the NR simulation is very expensive in terms of time and computing resources.

Therefore, we adopt the `parabolic approximation' to get the physical quantities of the marginally capturing hyperbolic orbit as we did in Paper I. It assumes that the radiated energy from the parabolic orbit is the same with that from the weakly hyperbolic orbit with the same pericenter, because their paths around the pericenter are almost the same. If the radiated energy from the parabolic orbit is obtained, it could be considered as the orbital energy of the marginally capturing hyperbolic orbit, since the orbital energy of the marginally capturing hyperbolic orbit would be the same with its radiated energy by GWs. For instance, if the radiated energy from the parabolic orbit is 1\% of total ADM (Arnowitt, Deser and Misner) energy, we can consider two BHs with the hyperbolic orbit whose orbital energy is 1\% of total ADM energy would be marginally captured because they would emit the same amount energy. With the parabolic approximation, we only need to calculate just one orbit---parabolic orbit for a given initial angular momentum. The cross sections of the capture in PN had been calculated based on this approximation \citep{quinlan87,quinlan89}, and we have also adopted it to NR results.

For the full general relativistic simulations, we have used the \textsc{Einstein Toolkit}\citep{loffler12}, and the simulation setup is basically the same with the Paper I except adding the spins to the BHs. In order to get more accurate initial conditions of spinning BHs, we have increased the number of spectral coefficients by more than twice the default values in \textsc{TwoPunctures} \citep{ansorg04}. For instance, we have taken 90 coefficients in the radial directions for high spin cases.

Initially, two BHs need to be positioned at a sufficient distance because they should have enough time before encounter in order not to be affected by the initial junk radiations. The initial separation of two BHs is fixed 60 M in geometrized units. The parabolic orbit is obtained by having total ADM mass be the same with the sum of ADM masses of each puncture \citep{ansorg04}. The initial orbital angular momentum is given as the initial separation multiplied by the initial momentum with perpendicular direction. The detailed description of getting the initial parabolic orbit is described in Paper I.

The symmetries in the simulation domain cannot be utilized if the directions of BH's spins are arbitrary. In addition, large simulation domain is necessary for our study because the orbit is elongated and GWs should be extracted at larger distances for the orbit with larger pericenter distances. Therefore, multipatch coordinate system (\textsc{Llama} code, \citealt{pollney11}) is adopted to reduce the computational resources. In the central domain where two BHs are located, we use the \textsc{Carpet} \citep{Schnetter03} module which adopts the adaptive mesh refinement (AMR). The finest grid size in the mesh refinement around the BH is determined by considering the size of the BH. Since the apparent horizon of the BH with larger spin magnitude is smaller, the smaller grid---the higher resolution is used for them. The innermost grid sizes are determined by the convergence test as described in the Appendix.

For the wave extraction, we have used \textsc{WeylScal4} and \textsc{Multipole} modules \citep{baker02}, which extract the Weyl scalar $\Psi_{4}$ and decompose into the spin-weighted spherical harmonics. In order to calculate the radiated energy, we have used the modes up to $l=8$ which are credible because the higher modes are dominated by the numerical noise \citep{pollney11}.

In NR simulation, a geometrized unit system is used, which takes the speed of light $c$ and the gravitational constant $G$ as unity. In this unit system, time, mass and energy have the same unit of length, and the angular momentum has a unit of length squared. We have presented our results with geometrized unit system. The physical quantities in SI unit system can be obtained by restoring the $c$ and $G$ values. For instance, the unit length and time of 10-10 M$_{\odot}$ BBHs are about 30 km and 0.1 ms, respectively.

\section{Resutls} \label{sec:results}


\subsection{Radiated Energy with BH spins}

Since we focus on the effect of spins to the radiation from encountering two BHs, various combinations of spin directions and their magnitudes should be considered in numerical simulations. This parameter space of spin configurations, however, is quite large, so it would be impossible to cover all cases. In order to learn about the effects of the spin on the two-body dynamics, we first take a look at the PN results. The GW energy flux for a system of two spinning compact objects with an arbitrary eccentricity is given in the following form \citep{blanchet06}
\begin{equation}
    {\cal F} = \frac{G^3}{c^5}\left\{ f_{\rm{NS}} +f_{\rm{SO}} +f_{\rm{SS}} +{\cal O} \left(\frac{1}{c^{6}}\right)\right\}
\end{equation}
at 2.5PN order. Here $f_{\rm NS}$ is the non-spin contribution, $f_{\rm SO}$ the spin-orbit coupling part, and $f_{\rm SS}$ the spin-spin contribution. Instead of using two spin angular momenta variables $\mathbf{S}_1$ and $\mathbf{S}_2$, one may equivalently use any other two combinations of these vectors. It turns out that the spin dependence in $f_{\rm SO}$ at 2.5PN order comes only through the following combinations of spin and orbital angular momenta:
\begin{equation}
S_l = \mathbf{\pmb{\ell}} \cdot \mathbf{S} = \mathbf{\pmb{\ell}} \cdot \left( \mathbf{S}_1 +\mathbf{S}_2 \right) \qquad   \&   \qquad \Sigma_l = \mathbf{\pmb{\ell}} \cdot \mathbf{\Sigma} = \mathbf{\pmb{\ell}} \cdot \left( \frac{M}{m_2}\mathbf{S}_2 -\frac{M}{m_1}\mathbf{S}_1 \right).
\end{equation}
Here $\mathbf{\pmb{\ell}} = \mathbf{L}/|\mathbf{L}|$ is the instantaneous Newtonian orbital angular momentum unit vector and $M=m_1+m_2$. It is interesting to notice that $\Sigma_l$ appears in $f_{\rm SO}$, at least, at 2.5PN order only with the multiplication of the mass difference $\delta m = m_1-m_2$, {\it i.e.,} $\sim \delta m \, \Sigma_l$. Consequently, for the case of equal masses, the spin contribution in the SO part depends only on the orbital angular momentum component of the sum of spins.

By using the dimension-less spin parameters $\pmb{\chi}_i$, we have $\mathbf{S}_i = m^2_i \pmb{\chi}_i$, and so $\Sigma_l = M \mathbf{\pmb{\ell}} \cdot \left( m_2 \pmb{\chi}_2 -m_1 \pmb{\chi}_1 \right)$. Defining
\begin{equation}
    \chi_{\mathrm{eff}}^{(\pm)} = \mathbf{\pmb{\chi}}_{\pm} \cdot \mathbf{\pmb{\ell}} = \left( \frac{m_2}{M} \mathbf{\pmb{\chi}}_2 \pm \frac{m_1}{M} \mathbf{\pmb{\chi}}_1 \right) \cdot \mathbf{\pmb{\ell}},
\end{equation}
we get
\begin{equation}
    \Sigma_l = M^2 \chi_{\mathrm{eff}}^{(-)}, \qquad S_l =
    \frac{M^2}{2} \chi_{\mathrm{eff}}^{(+)} -\frac{M}{2} \delta m \chi_{\mathrm{eff}}^{(-)}.
\end{equation}
Therefore, the spin effect up to 2.5PN order on the radiated energy by GWs due to the SO coupling can be described completely by two parameters, {\it e.g.,}
\begin{equation}
   {\cal F}_{\rm SO} = A M\chi_{\mathrm{eff}}^{(+)} +B \delta m \chi_{\mathrm{eff}}^{(-)} +\cdots.
\end{equation}
This feature is consistent with the simple expectation of invariance under the flipping of the orbital plane (i.e., $\mathbf{\pmb{\ell}} \rightarrow -\mathbf{\pmb{\ell}}$, $\mathbf{\pmb{\chi}}_{1, 2} \rightarrow -\mathbf{\pmb{\chi}}_{1, 2}$). The fact that $\chi_{\mathrm{eff}}^{(-)}$ always comes with $\delta m$ may also be understood because the systems should physically be the same under the exchange of two objects $(1 \leftrightarrow 2)$ for which $\mathbf{\pmb{\chi}}_{+} \rightarrow + \mathbf{\pmb{\chi}}_{+}$, $\mathbf{\pmb{\chi}}_{-} \rightarrow - \mathbf{\pmb{\chi}}_{-}$, $\mathbf{\pmb{\ell}} \rightarrow +\mathbf{\pmb{\ell}}$, and $\delta m \rightarrow -\delta m$, yielding
$\chi_{\mathrm{eff}}^{(+)} \rightarrow +\chi_{\mathrm{eff}}^{(+)}$ and $\chi_{\mathrm{eff}}^{(-)} \rightarrow -\chi_{\mathrm{eff}}^{(-)}$. Note also that the ranges of these two effective spin parameters are both $-1 \leq \chi_{\mathrm{eff}}^{(\pm)} \leq 1$. Identical spins $\mathbf{\pmb{\chi}}_1 = \mathbf{\pmb{\chi}}_2 = \mathbf{\pmb{\chi}}$ aligned along the orbital angular momentum give $\chi_{\mathrm{eff}}^{(+)} = \mathbf{\pmb{\chi}} \cdot \mathbf{\pmb{\ell}}: 0 \sim 1$ and $\chi_{\mathrm{eff}}^{(-)} = (-\delta m/M) \mathbf{\pmb{\chi}} \cdot \mathbf{\pmb{\ell}}$. Anti-aligned spins $\mathbf{\pmb{\chi}}_1 = \mathbf{\pmb{\chi}}_2 = -\mathbf{\pmb{\chi}}$, on the other hand, give $\chi_{\mathrm{eff}}^{(+)} = -\mathbf{\pmb{\chi}} \cdot \mathbf{\pmb{\ell}} : -1 \sim 0 $ and $\chi_{\mathrm{eff}}^{(-)} = (\delta m/M) \mathbf{\pmb{\chi}} \cdot \mathbf{\pmb{\ell}}$.

The spin-spin part at 2PN order, on the other hand, depends on the components of spins on the orbital plane and $\mathbf{S}_1 \cdot \mathbf{S}_2$ \citep{kidder95,cho98}. Thus, this contribution could be more sensitive to various configurations of spins than the spin-orbit part. Its overall effect on the radiated energy in total, however, turns out to be negligible as can be seen in our NR simulation results below. This behavior may follow because the spin-spin contribution appears at PN order higher than the leading order in the spin-orbit contributions.

In this paper, we consider the encounters of BHs with identical masses only, for which $\delta m =0$ and so terms with $\chi_{\mathrm{eff}}^{(-)}$ do not contribute. Therefore, $\chi_{\mathrm{eff}}^{(+)}$ is expected to be the main parameter that determines the GW radiations of the two equal-mass BHs even in full GR, and it is indeed used to design the parameter space at the present paper. Hereafter, we will simply denote $\chi_{\mathrm{eff}}^{(+)}$ as $\chi_{\mathrm{eff}}$.

\begin{figure}[ht!]
    \plotone{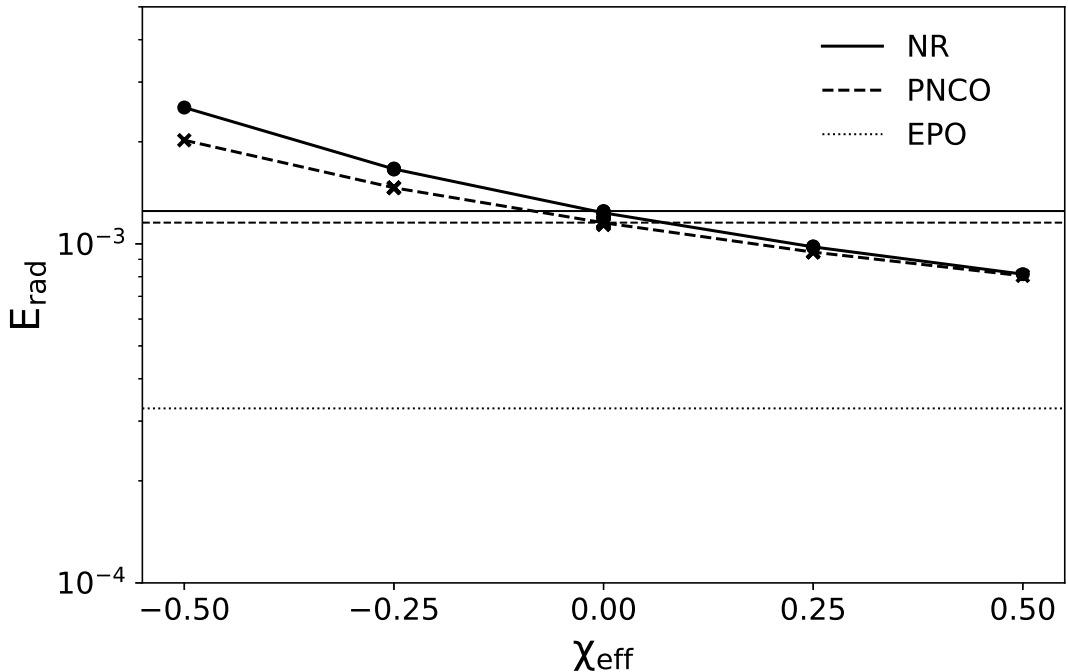}
    \caption{The radiated energies with different spin directions. The mass ratio, initial orbital angular momentum and the spin magnitudes are fixed as $m_{1}/m_{2}=1$, $\Linit=1.11$ and $|\pmb{\chi}_{1}|=|\pmb{\chi}_{2}|=0.5$, respectively, but the spin directions are changed by 90 degrees along the x, y and z axes in order to have different $\chi_{\mathrm{eff}}$. The cases with the same \chieff~are almost overlapped. The horizontal lines are from non-spinning cases (solid: NR, dashed: PNCO, dotted: EPO) \label{sp_direct}}
\end{figure}

In order to verify our conjecture on the effect of spins, we have fixed the spin magnitudes as $|\pmb{\chi}_{1}|=|\pmb{\chi}_{2}|=0.5$ for both two BHs and changed the directions by 90 degrees on x, y and z directions. In that case, there are 21 combinations except the duplications by rotating 180 degrees. The initial orbital angular momenta are fixed as $\Linit=1.11$ which gives fly-by orbits for all spin combinations. Although the total angular momenta are different from each other because of the spin angular momenta, we fix the initial orbital angular momenta for the comparison with the non-spinning cases. Note that the initial energies in all cases are also the same because all initial orbits are parabolic.

Figure \ref{sp_direct} shows the radiated energies with respect to the effective spin parameter $\chi_{\mathrm{eff}}$. \chieff$=0.5$ represents the spins of both BHs are aligned and \chieff$=-0.5$ does the spins of both BHs are anti-aligned with the orbital angular momentum. \chieff$=0.25$ or \chieff$=-0.25$ represent that one BH has aligned or anti-aligned spin with respect to the orbital angular momentum, and the other BH has the spin with perpendicular to the orbital angular momentum. \chieff$=0$ means both spins are perpendicular to the orbital angular momentum, or one is aligned and the other is anti-aligned with respect to the orbital angular momentum direction.

As expected from PN results, radiated energy depends on the \chieff~monotonically: for a given initial orbital angular momentum, smaller \chieff~case radiates more than larger \chieff. The cases of anti-aligned spins with the orbital angular momentum emit more energy, while the aligned cases do less. The points at the same \chieff~are almost overlapped. For instance, when the both spin directions are perpendicular to the orbital direction, or one is aligned and the other is anti-aligned (\chieff=0), the amount of GW radiation is almost the same with non-spinning case within the 3\% differences. Twelve cases with different spin directions are overlapped on \chieff=0 including non-spinning cases, and four cases are overlapped on each  $\chi_{\rm eff}=-0.25$ or 0.25, respectively.

\begin{figure}[ht!]
    \plotone{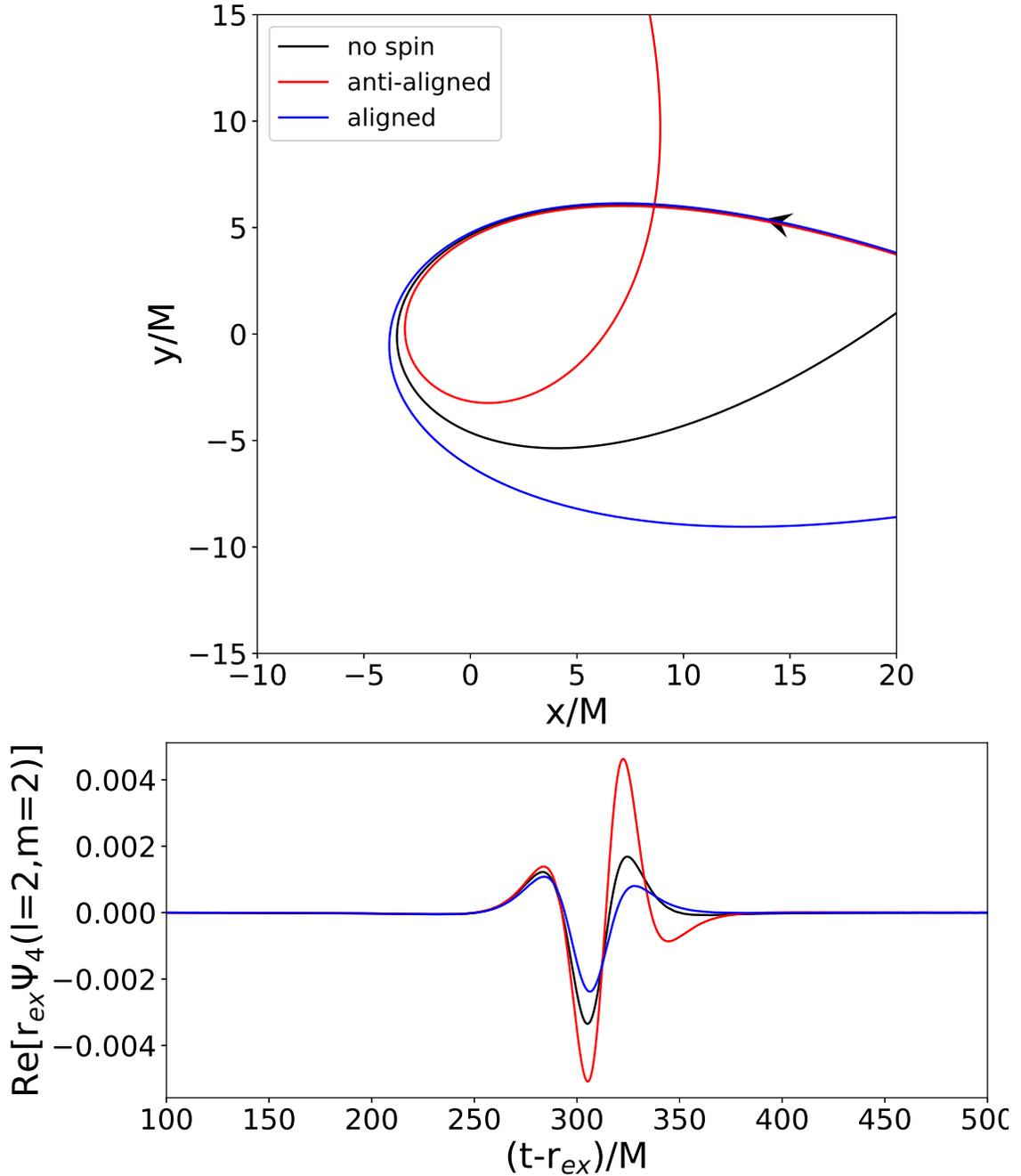}
    \caption{The orbits and the Weyl scalars $\Psi_{4}$ are presented for different spin directions. The initial orbital angular momenta are fixed as $\Linit=1.11$ and the spin directions are aligned or anti-aligned with the orbital angular momentum directions. In the upper panel, we draw just one BH's orbit, the other is in the opposite side of the origin. An arrow indicates the orbital direction. The spin magnitudes are fixed as $|\pmb{\chi}_{1}|=|\pmb{\chi}_{2}|=0.5$. The Weyl scalars are multiplied by the wave extraction radii ($r_{\rm ex}$) and time is also subtracted by $r_{\rm ex}$ in order to remove their dependencies. \label{sp_orbit_wave}}
\end{figure}

These results can be also explained analytically from a perspective of the orbits. The spin-orbit coupling term in PN equation of motion (equation above (A.3) in \citealt{levin11}) is as follow
\begin{equation}
    \mathbf{a}_{\rm{PN-SO}}=\frac{1}{x^3}\left\{ 6\frac{\mathbf{\hat{n}}}{x}\tilde{\mathbf{L}}_{N}\cdot(\pmb{\mathscr{S}}+\pmb{\xi}) - \mathbf{v}\times(4\pmb{\mathscr{S}}+3\pmb{\xi}) + 3\dot{x}\mathbf{\hat{n}}\times(2\pmb{\mathscr{S}}+\pmb{\xi}) \right\}~,
\end{equation}
where $\pmb{\mathscr{S}}=(m_{1}^{2}\bm{\chi_{1}}+m_{2}^{2}\bm{\chi_{2}})/(m_{1}+m_{2})^{2}$ and $\pmb{\xi}=m_{1}m_{2}(\pmb{\chi_{1}}+\pmb{\chi_{2}})/(m_{1}+m_{2})^2$ are the spin parameters, $x$, $\mathbf{v}$ and $\mathbf{\hat{n}}$ are relative distance, velocity and normalized position vector, respectively, $\tilde{\mathbf{L}}_{N}$ is reduced Newtonian orbital angular momentum, and $\dot{x}=\mathbf{\hat{n}}\cdot\mathbf{v}$ is relative velocity in the position vector direction.

In the case of equal masses, $\pmb{\mathscr{S}}$ and $\pmb{\xi}$ are the same, thus the first term is directly connected to the \chieff. The third term is negligible around the pericenter because $\dot{x}\sim 0$. The second term can be considered in three cases. First when $\mathbf{v}\parallel\pmb{\mathscr{S}}$, the second term vanishes. Second case is when $\mathbf{v}\perp\pmb{\mathscr{S}}$ and $\pmb{\mathscr{S}}$ is not in the orbital plane, the direction of the second term is along the $\mathbf{\hat{n}}$ direction. $\pmb{\mathscr{S}}$ is zero when \chieff$=0$. When the spins are aligned to the orbital angular momentum (\chieff$=0.25$ or $0.5$), the second term is parallel to the $\mathbf{\hat{n}}$ direction, making the pericenter distance larger. On the other hand, the second term has the opposite direction to the $\mathbf{\hat{n}}$ when the spins are anti-aligned to the orbital angular momentum (\chieff$=-0.25$ or $-0.5$), making the pericenter distance smaller. Third case is when $\mathbf{v}\perp\pmb{\mathscr{S}}$ and $\pmb{\mathscr{S}}$ is in the orbital plane, so that the second term to be in perpendicular direction to the orbital plane, thus it makes relatively small difference in the pericenter distance. Thus the pericenter distances as well as the orbits can be different depending on the spin directions, which leads the differences in radiated energies as seen in Figure \ref{sp_direct}.

\begin{figure}[ht!]
    \plotone{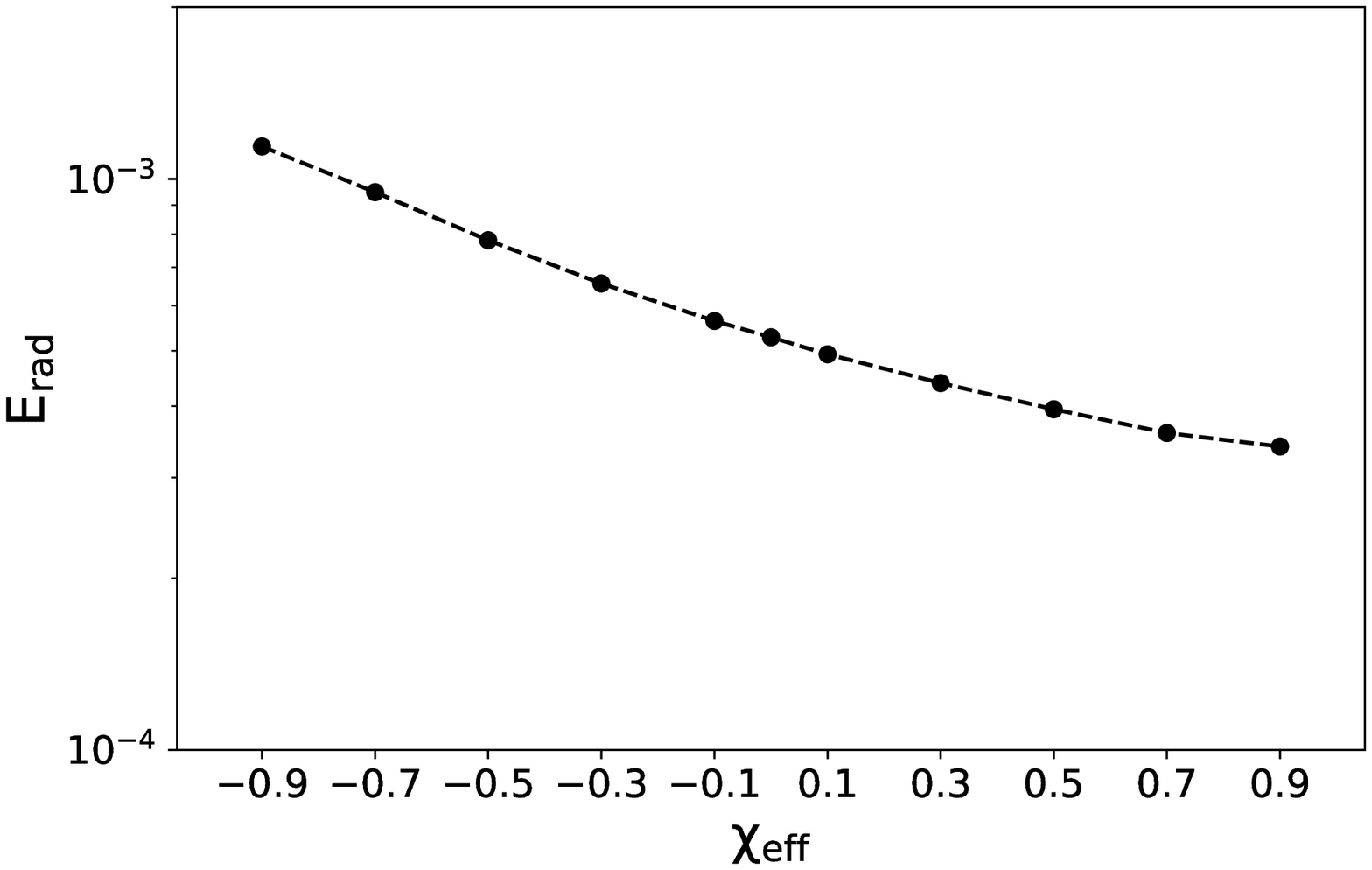}
    \caption{Radiated energies for various spin magnitudes. The initial orbital angular momenta are fixed as $\Linit=1.2$. Both BH spin directions are aligned or anti-aligned with respect to the orbital angular momentum direction and both spin magnitudes are the same. The spin magnitudes are changed at intervals of 0.2. \label{sp_mag}}
\end{figure}

Figure \ref{sp_orbit_wave} shows the orbits and waves when the initial orbital angular momenta are fixed as $\Linit=1.11$ and both spin directions are anti-aligned or aligned with the orbital angular momentum direction. The anti-aligned spin case gives tightly wound orbit, while the aligned spin case shows loosely wound orbit. Therefore, the radiated energies are different for different combination of spin directions---stronger for anti-aligned case and weaker for aligned. Some cases which have the other combination of spin directions include the z-directional orbits, not confined in xy plane. However, they are not our concern because their pericenter distances are almost the same with the same \chieff~cases. Thus they are not special in the aspect of the radiated energy shown in Figure~\ref{sp_direct}.

The simple PN results are also drawn for comparison on Figure~\ref{sp_direct}. Two different orbits---exactly parabolic orbit (EPO) in Newtonian, and PN corrected orbit (PNCO) using the PN equation of motion upto 3.5 PN including the spin-orbit and spin-spin effects \citep{kidder95,tagoshi01,will05,faye06,levin11,blanchet14}---are considered, and the radiated energies are calculated along the orbits using the quadrupole formula \citep{peters63,peters64}. EPO cases always underestimate the amount of GW radiations, even for the aligned spin case \chieff$=0.5$ which gives the least GW radiations. PNCO also underestimates the radiated energy for this initial orbital angular momentum, especially for the anti-aligned spin case, but gives much better agreements with the NR results.

Then, we have changed the spin magnitudes with fixed spin directions (Figure~\ref{sp_mag}). Both BH spin directions are aligned or anti-aligned simultaneously to the orbital angular momentum, and spin magnitudes are the same with each other and changed at intervals of 0.2. The positive \chieff~represents that the spin directions are aligned to the orbital angular momentum and the negative \chieff~represents the anti-aligned spins to the orbital angular momentum. As expected, the case of large spin magnitudes gives large deviation from the non-spinning case.

\begin{figure}[ht!]
    \plotone{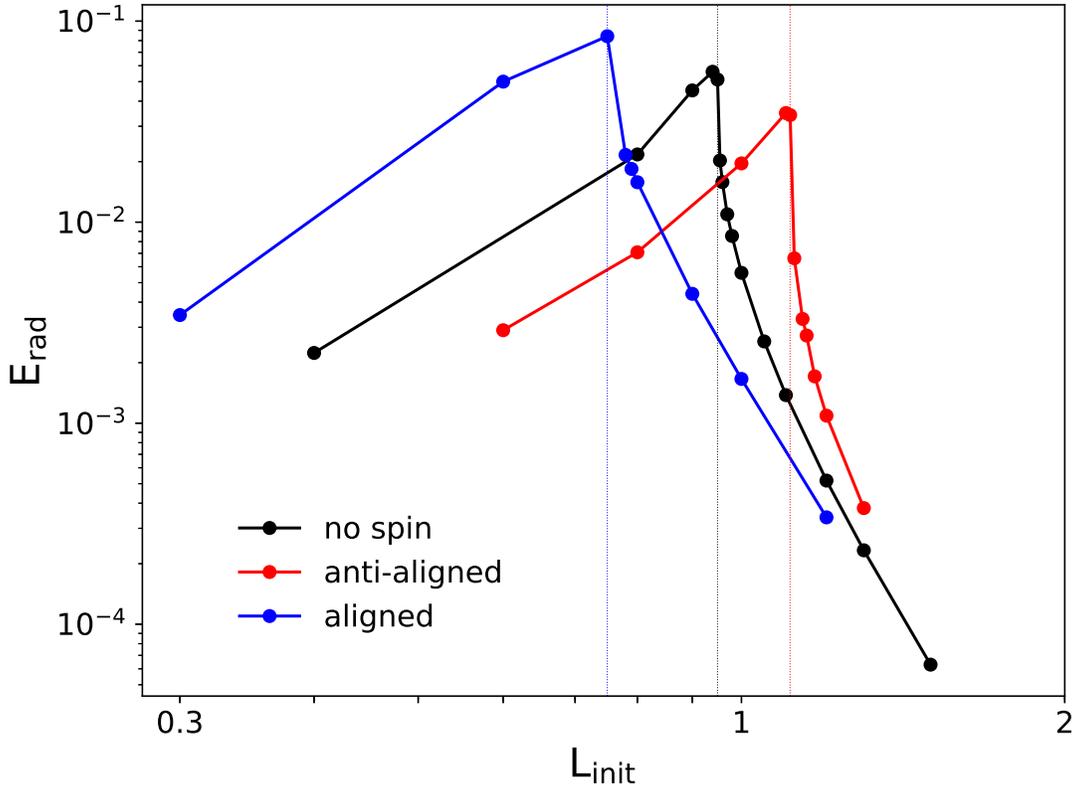}
    \caption{The radiated energies when both BHs have aligned or anti-aligned spins ($|\pmb{\chi}_{1}|=|\pmb{\chi}_{2}|=0.9$) relative to the orbital angular momentum. Note that here $\Linit$ is the initial orbital angular momentum, not the total angular momentum including the spin angular momentum.  \label{radEL}}
\end{figure}

\subsection{Radiated Energy with initial orbital angular momentum}

Next, we have imposed the higher spins on the BHs for the aligned (\chieff$=0.9$) and anti-aligned (\chieff$=-0.9$) spin directions. Since the BH whose spin is close to extremal spin requires different approach \citep{lovelace11,lovelace12,lovelace15,scheel15,ruchlin17}, we chose the spin magnitudes for both BHs as 0.9 which are reasonably available in our numerical setup.

\begin{figure}[ht!]
    \plotone{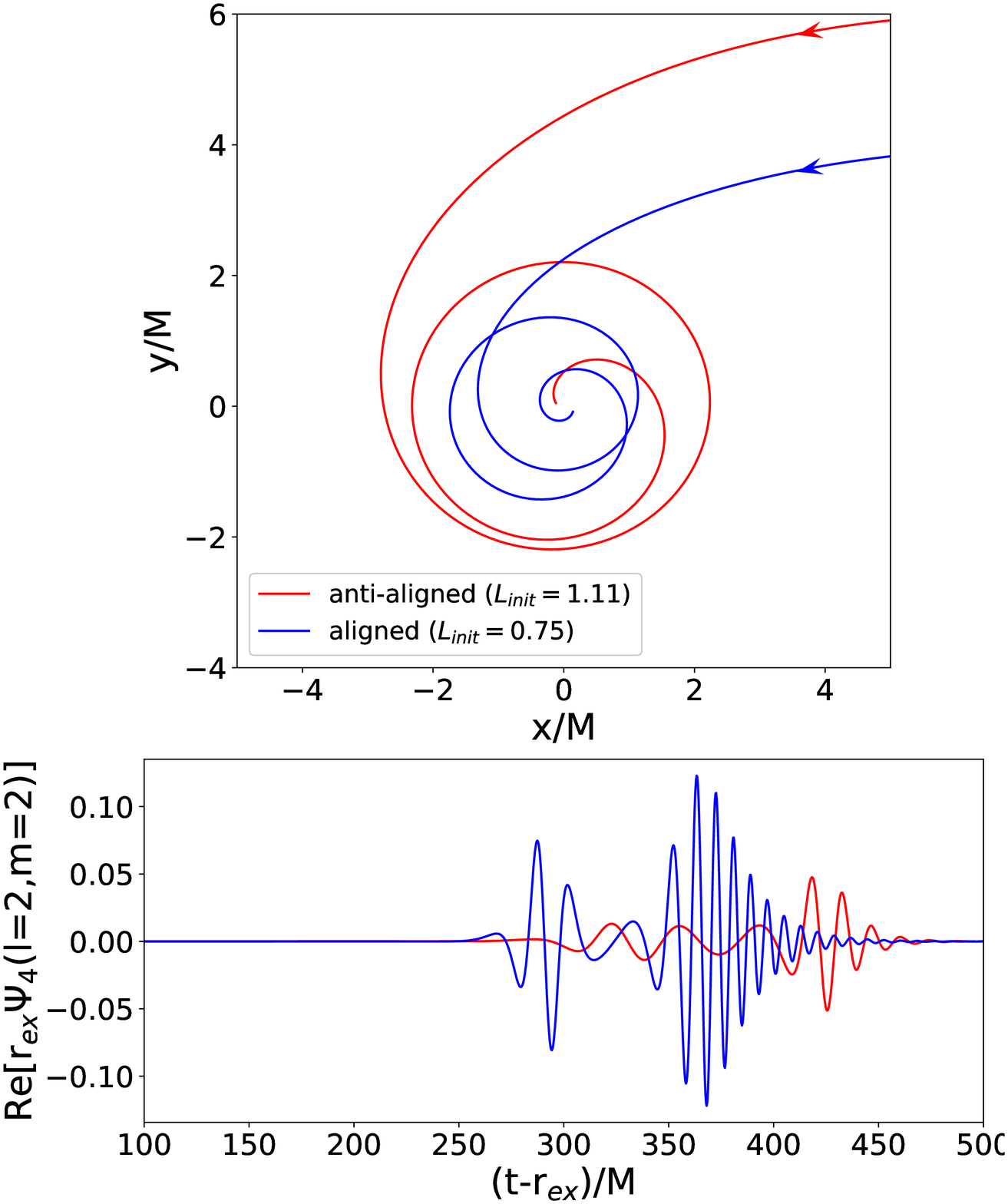}
    \caption{The orbits and GWs at the peak point where the most GWs are emitted. The aligned case has more tightly wound and whirly orbit. The common horizon is formed at around $t=350$ M for aligned case and $t=410$ M for anti-aligned case. \label{sp_peak_orbit}}
\end{figure}

Figure \ref{radEL} shows the radiated energies for different initial orbital angular momenta ($\Linit$). As the $\Linit$~decreases, the pericenter distance becomes smaller and the amount of radiated energy becomes larger, and eventually two BHs merge directly. The orbits of the right parts of dotted lines are fly-by, while those of the left parts including peak points are the direct merging orbits. At the fixed orbital angular momentum in the fly-by orbit, the anti-aligned spin cases emit more energies. This can be understood from the orbits shown in Figure \ref{sp_orbit_wave}. The orbits of anti-aligned spins give shorter pericenter distances. In fact, their total angular momenta are smaller than those of the non-spinning cases due to the spin angular momenta. On the other hand, the aligned spins give less GW radiations in the fly-by orbits at the given initial orbital angular momentum, because their pericenters are more distant.

 By contrast, the aligned spin cases emit more GWs in direct merging orbits, because their orbits are more tightly wound and whirly before the merging (Figure~\ref{sp_peak_orbit}). This is consistent with the results of \citet{campanelli06} for the quasi-circular orbits. Even after the merging, the merged BH emits more GWs through the long ringdown phase. Thus, we should note that the fly-by and the direct merging orbits give opposite trends in GW radiations. In next section, we have just used the results of the fly-by orbits to calculate the critical impact parameter because only those are valid for the parabolic approximation.

\begin{figure}[ht!]
    \plotone{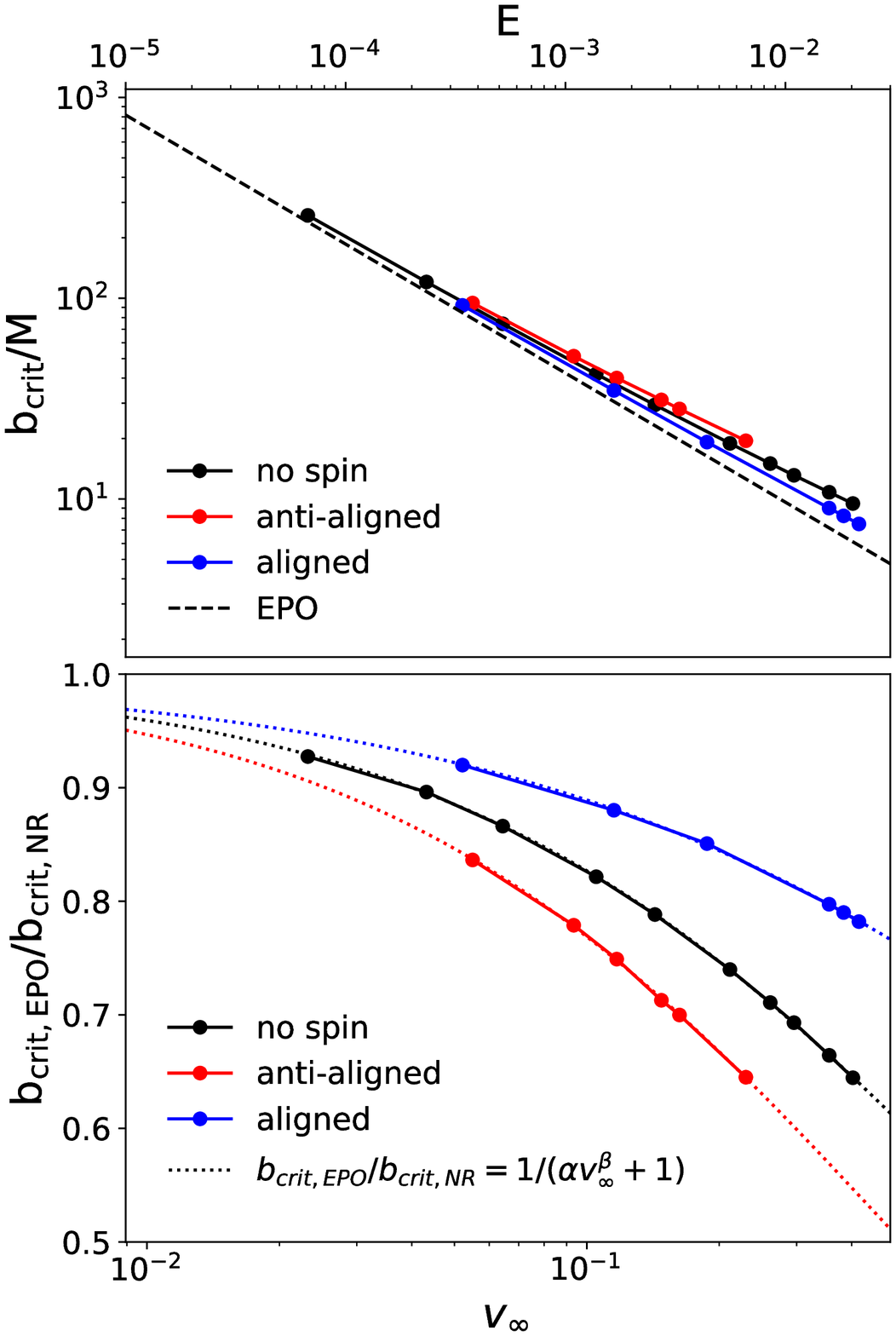}
    \caption{The critical impact parameters ($\bcrit$) with respect to the relative velocity at infinite distance ($v_{\infty}$). The spins of two BHs are aligned or anti-aligned to the direction of the orbital angular momentum. The spin magnitudes are fixed as $|\pmb{\chi}_{1}|=|\pmb{\chi}_{2}|=0.9$. Low panel shows the relative size of the impact parameters of spinning BHs between NR and EPO.}
    \label{imp_ev}
\end{figure}


\subsection{Critical Impact Parameters for Capture} \label{subsec:crtimpt}

The critical impact parameter $\bcrit$ for GW capture can be calculated under the parabolic approximation as follows. The orbital angular momentum of the marginally capturing hyperbolic orbit is given as $L=b_{\rm crit}\mu v_{\infty}$ where $\mu$ is the reduced mass of two BHs and $v_{\infty}=\sqrt{2E/\mu}$ is the relative velocity at infinite distance which can be expressed by the orbital energy $E$. Since the capture occurs when the radiated energy is greater than the orbital energy $E$, the critical impact parameter for capture $b_{\rm crit}$ can be obtained by
\begin{equation}\label{eq:bcrit}
    b_{\rm crit}=\frac{L}{\mu v_{\infty}}=\frac{L_{\rm init}}{\sqrt{2\mu E_{\rm rad}}}~.
\end{equation}
Especially in the EPO, which adopts Newtonian parabolic orbit, the critical impact parameter can be expressed analytically as
\begin{equation}
   b_{\rm crit,EPO}=C\vinf^{-9/7}~,
\end{equation}
where $C=\left(340\pi m_{1}m_{2}(m_1+m_2)^5/3\right)^{1/7}$ is given by Eq. (8) in Paper I for non-spinning BHs.

Figure \ref{imp_ev} shows the behavior of $\bcrit$ as the initial velocity at infinity $v_{\infty}$ varies. Here $v_{\infty}$ for a weakly hyperbolic orbit is calculated by the radiated energy obtained from the parabolic approximation. The capturing cross section for encountering spinning BHs is given simply by $\sigma=\pi \bcrit^{2}$. The spin effect on GW capturing processes grows as the relative velocity of encountering BHs at infinity increases. In the region of very high velocities, critical impact parameters are split, depending on the spin directions. The anti-aligned spin case gives larger impact parameter, while the aligned spin case gives smaller one. The cases of any other spin configurations will be located between these two extreme cases. Note that all cases do not give smaller impact parameter than the EPO's. Regardless of the spin directions, the numerical relativity gives larger cross sections than EPO.

As can be seen in Figure \ref{imp_ev}, the critical impact parameters from NR have been fitted by using non-linear least squares. With the analytic expression for $b_{\rm crit,EPO}$ above, these fitting formulas become
\begin{equation}\label{eq:imp_fit}
   b_{\rm crit,NR} = C \vinf^{-9/7} (\alpha \, \vinf^{\beta}+1)~.
\end{equation}
 Here the pairs ($\alpha$,$\beta$) are (1.03, 0.69), (1.61, 0.73) and (0.46, 0.56) for zero, anti-aligned and aligned spins, respectively, within about 0.3\% errors. These factors can be used for the correction of PN results in the highly energetic encounters. At the low $\vinf$ limit, $\bcrit\propto\vinf^{-9/7}$ as expected from PN, but at the high $\vinf$ limit where $\vinf$ approaches to unity---speed of light, the critical impact parameter could become larger than PN value by a factor of $\sim1.5$ for anti-aligned spin cases as shown in the lower panel of Figure \ref{imp_ev}. If $v_{\infty}$ is larger than $\sim0.1$ ($\bcrit\sim50$ M), the EPO could underestimate the critical impact parameter by less than 80\% for the anti-aligned spin cases. But in the range of $v_{\infty}<0.01$, the EPO gives a good estimation for critical impact parameters more than 95\%.


\section{Discussions} \label{sec:discuss}

We have calculated the critical cross sections for GW capture of two BHs in an unbound orbit by carrying out the parabolic orbit simulations. In fact all the orbits we have simulated are bound orbits because they are initially parabolic. But note that we have assumed the parabolic approximation, which regards the radiated energy from the parabolic orbit as the orbital energy of the hyperbolic orbit that can be marginally captured by each other. The radiated energy from just one-time passage is used for the calculation of critical cross section. The directly merging orbit cannot be used for that.

However, the one-time passage in the fly-by orbit is not easy to define. If the apocenter is sufficiently distant, the one-time passage would be easily defined. But at the initial orbital angular momentum around the boundary between direct merging and fly-by orbits (Figure~\ref{sp_peak_orbit}), the orbits of two BHs swirl tightly around each other, especially for the aligned spin cases. In those cases, it is hard to distinguish the one-time passage from the orbit. Thus, we have considered those cases as the direct merging orbits and excluded from the calculations of the critical impact parameters.

The velocities at infinity $\vinf$ in Figure \ref{imp_ev} seem unrealistically high from an astrophysical point of view. Typical central velocity dispersion of globular clusters, Galactic center, and M31 are about 10, 75, and 160 km s$^{-1}$, respectively \citep{harris96,gebhardt00}. Even in the center of the massive elliptical galaxies, the velocity dispersion is about 300 km s$^{-1}$ \citep{duncan99}. But we should note that the velocities at infinity $\vinf$ in Figure \ref{imp_ev} are just the maximum values that can be captured for given impact parameters. Whether we need to consider full relativistic effects should be closely related to the closest distance between the BHs during the encounters.

Let us estimate what percentage of GW captures needs the relativistic orbit corrections in astrophysical situation. According to the eq. (2) in \citet{samsing19}, the maximum pericenter distance for GW capture of equal mass BHs is given as $r_{\rm p,max}\approx1.8v_{\infty}^{-4/7}$ M.  Meanwhile, the pericenter distance of the $\Linit=1.5$ with no spin case, which has the weakest encountering orbit in our simulations, is about 15 M. This is very small distance considering that the radius of the inner stable circular orbit of a test particle around a BH is 6 M, and the deviation of impact parameter of EPO is more than 5\% (Figure~\ref{imp_ev}) for this case. If we take this as a criterion of whether the relativistic correction of the orbit is necessary, about $(15/1.8) v_{\infty}^{4/7}$ of GW captures needs the relativistic correction in orbit, since the pericenter distances have a uniform distribution \citep{hong15} for encounters in Newtonian regime. For example, the orbits of more than 10\% of GW captures need the relativistic correction if we assume the relative velocity of BHs are 150 km s$^{-1}$.

The fraction of direct merging cases can also be calculated roughly by using simple Newtonian dynamics. The impact parameter $b$ can be expressed with the pericenter distance $r_{\rm p}$ as follow from the orbital angular momentum and energy at the pericenter and the infinite distance,
\begin{equation}\label{eq:b_rp}
    b=r_{\rm p} \sqrt{1+\frac{2G(m_{1}+m_{2})}{r_{p}v_{\infty}^{2}}}~.
\end{equation}
We can calculate the pericenter distance corresponding to a certain orbital angular momentum using Eq. (\ref{eq:bcrit}) and (\ref{eq:b_rp}) as follow
\begin{equation}\label{eq:rp_b_L}
    r_{\rm p}=-\frac{1}{\vinf^{2}}\left[1-\sqrt{1+\frac{L^2}{\mu^2}\vinf^{2}} \right]~,
\end{equation}
which can be approximated to $r_{\rm p}\approx L^{2}/2\mu^{2}$ when $\vinf$ is small. According to our simulation results, two BHs have direct merging orbits when their initial angular momenta are 0.75 for aligned spins and 1.11 for anti-aligned spins when the spin magnitudes are 0.9 (Figure~\ref{radEL}). The corresponding pericenter distances are about $r_{\rm p}=4.5$ M and 10 M, which means that about 3-7\% of GW captured BHs can have direct merging orbit in the cluster with the velocity dispersion of 150 km s$^{-1}$, considering the maximum pericenter distance $r_{\rm p,max}$ for GW capture and the uniform distribution of $r_{\rm p}$ again. The fraction of direct merging cases in GW capture increases proportionally to $\vinf^{4/7}$.

The waveforms produced during the GW capture have interesting features. From the highly eccentric orbits after GW capture, we can expect the sporadic burst-like waves because strong GWs are emitted only when they pass by pericenter \citep[e.g.,][]{hong15}. For the direct merging orbits, merger and ringdown without inspiral phase are expected (Figure 2 in \citealt{bae17}). The combination of them are also possible if they have zoom and whirl orbits (Figure~\ref{sp_peak_orbit}), even though its condition is very restrictive. New type of GW templates will be necessary to search for those kind of GWs.

In this study, BH's spin magnitudes of $|\pmb{\chi}_{1}|=|\pmb{\chi}_{2}|=0.9$ or less are considered due to the limitation of Bowen-York initial condition \citep{bowen80,cook90,dain02}. It would have been better if we could test the higher spin cases, but instead, we presented the trend with the spin magnitudes in Figure~\ref{sp_mag} from which we can guess the higher spin cases. More highly spinning case would lead to the larger spin effects, but even maximal spinning case is not likely to be much different from $|\pmb{\chi}_{1}|=|\pmb{\chi}_{2}|=0.9$ cases, considering their trends.

The parabolic approximation we have adopted in this study has been used in previous works \citep{quinlan87,quinlan89,hansen14,bae17} because of the advantage of being able to obtain the initial conditions easily. But as checked in Paper I, real marginally capturing hyperbolic orbit without the parabolic approximation is expected to emit more energy in the extremely close encounter, which gives larger critical impact parameter. In addition, the PN approximation for the leading order spin-orbit part of the energy emitted shows that cases of unequal mass spinning BHs give non-vanishing contributions associated with the spin parameter $\chi_{\mathrm{eff}}^{(-)}$. It will be of interest to see in full GR how much the degeneracy in spin effects breaks up due to unequal masses. We will study those cases in near future.


\section{Acknowledgments}
This work was supported in part by the National Research Foundation of Korea (NRF) grant funded by the Korea government (MEST) (No. 2019R1A2C2006787), National Institute for Mathematical Sciences (NIMS) funded by Ministry of Science, ICT (B19720000), and the National Supercomputing Center with supercomputing resources including technical supports (KSC-2018-CHA-0027). GK was supported by the Korea Institute of Science and Technology Information (K-19-L02-C07, K-20-L02-C09). We would like to thank Jinho Kim, Chan Park, Sang-Hyeon Ahn, Jongsuk Hong and Sang Hoon Oh for useful discussions.


\appendix

\section{Convergence Test} \label{sec:convg}

\begin{figure}[ht!]
    \plotone{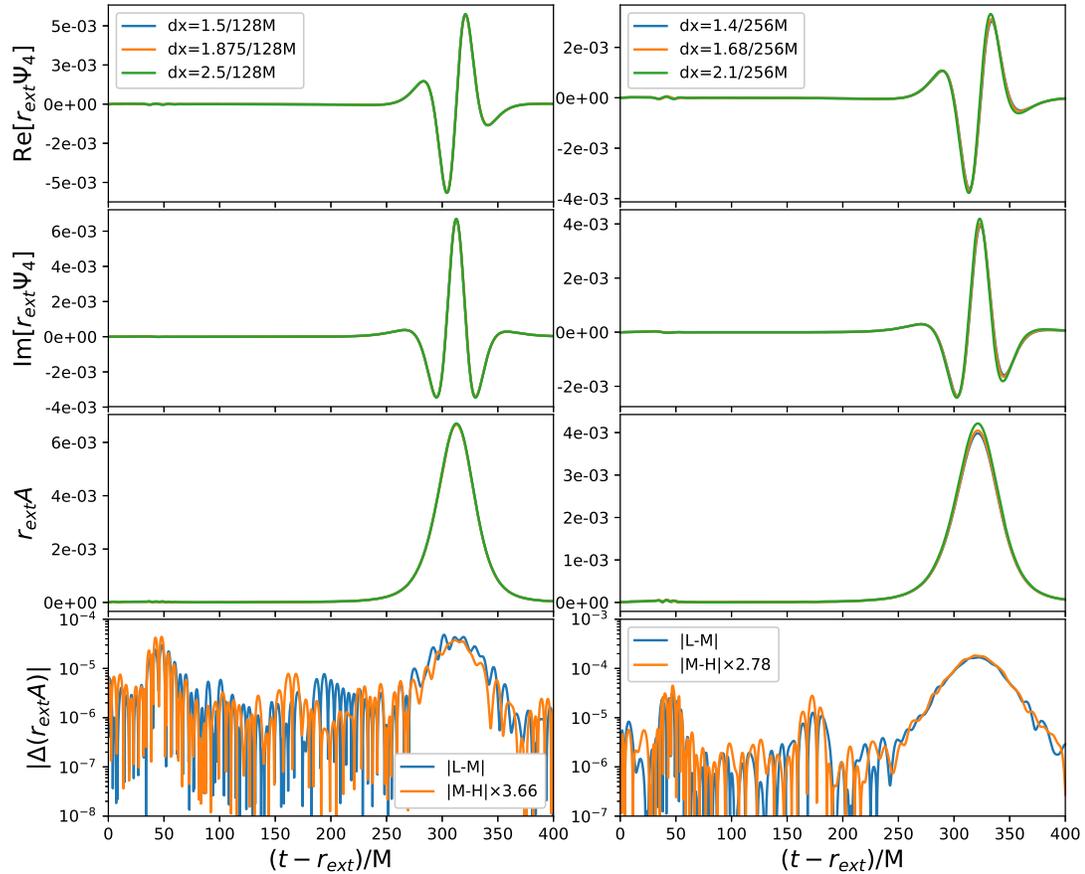}
    \caption{The comparison of the $\Psi_{4}$ and their amplitudes from different resolutions when $(l,m)=(2,2)$. The lowest panels show the absolute values of the differences of the $\Psi_{4}(l=2,m=2)$ between the different resolutions. The notation `L', `M' and `H' mean low, medium and high resolutions. They have nearly 4th order convergence.}
    \label{fig:convg}
\end{figure}

We have checked the convergence of our simulations. Two representative spin magnitudes, $|\pmb{\chi}_{1}|=|\pmb{\chi}_{2}|=0.5$ and 0.9 are selected, and each of them is used in two different cases in the test. First case has the spin magnitudes of $|\pmb{\chi}_{1}|=|\pmb{\chi}_{2}|=0.5$ for both BHs and their initial orbital angular momentum is $\Linit=1.1$. Second case has the spins of $|\pmb{\chi}_{1}|=|\pmb{\chi}_{2}|=0.9$ with $\Linit=1.17$. In both cases, the spin directions of two BHs are anti-aligned to the orbital angular momentum direction.

For each case, three different resolutions are tested. For the first case the smallest grid sizes in the AMR around the BH are dx$=1.5/128$ M, $1.875/128$ M and $2.5/128$ M, and for the second case dx$=1.4/256$ M, $1.68/256$ M and $2.1/256$ M are used.

The upper two rows in Figure~\ref{fig:convg} show the comparison of the real and imaginary parts of the Weyl scalar $\Psi_{4}$ with $l=2$ and $m=2$ mode, and third rows are amplitudes from them. The Weyl scalars are extracted at 200 M, but in the plot the Weyl scalars and amplitudes are multiplied by the extraction radius so as to see without dependence on that. The time is also subtracted by the extraction radius in order to see without the retardation.

The last rows show the absolute values of the differences between the amplitudes among the three different resolutions. At around the peak points, the difference between low and mid resolution is almost overlapped by the difference between mid and high resolution multiplied by some factors. These factors---3.66 and 2.78 in the plot---represent the 4th order convergences (see eq. (1) in paper I), which means that our simulations have almost 4th order convergences.

Basically, we have adopted the 8th order finite difference methods in the simulations but the lower order treatments in the mesh refinement or the multipatch make the overall convergences lower \citep{pollney11}. In the main text, we have used the resolutions which are between low and high resolution in this convergence test as dx=1.0/64 M and dx=1.8/256 M for $|\pmb{\chi}_{1}|=|\pmb{\chi}_{2}|=0.5$ and 0.9, respectively.

\bibliography{GRcap_spin}{}
\bibliographystyle{aasjournal}

\end{document}